SHORT COMMUNICATION

# The Consistency of Modeled and Observed Temperature Trends in the Tropical Troposphere: A Comment on Santer et al (2008)

Stephen McIntyre
Climate Audit
Toronto, Ontario

Ross McKitrick
University of Guelph
Guelph Ontario

**ABSTRACT**
Santer et al (2008) (S08) compared climate models and observations in the tropical troposphere and reported that "there is no longer a serious discrepancy between modeled and observed trends in tropical lapse rates." They found no statistically significant differences between modeled (ensemble mean) trends and observed trends at the T2LT and T2 layers, and they found no significant difference between observed and modeled surface-minus-troposphere lapse rates. However they only used data over the 1979-1999 period. Using the S08 methodology on up-to-date data, we find a statistically significant discrepancy between observations and models with respect to trends in the UAH data, as well as lapse rate trends comparing either RSS or UAH to the HADCRUT3v land-ocean surface trend.

The Consistency of Modeled and Observed Temperature Trends in the Tropical Troposphere: A Comment on Santer et al (2008)

**Introduction**

Santer et al (2008) ("S08") analyzed what they termed a "potential inconsistency" between modelled and observed trends in tropical lapse rates using "new observational estimates of surface and tropospheric temperature changes", concluding that "there is no longer a serious discrepancy between modelled and observed trends in tropical lapse rates". Their analyses involved both satellite and radiosonde data. Our comment pertains to the satellite data, which has been a source of ongoing controversy and which is worth examining in its own right. In respect to the satellite data, S08 carried out separate analyses comparing observed trends to the mean of an ensemble of 49 climate model runs and to the overall population of models. We restrict our comment to a re-assessment of their conclusions that were based on results involving the model ensemble mean. Similar examination of conclusions for the individual model runs would also be worthwhile.

S08 placed considerable importance on results for the "new" RSS satellite data set, of which they stated that "the surface warming is invariably amplified in the tropical troposphere, consistent with model results." For the UAH satellite data set, they reported both that the observed tropical lapse rate trends are not significantly different from models and that there was no statistically significant difference between the trends of the model ensemble mean and observations at the two tropospheric levels (T2LT, T2). These results have been relied on several recent assessment reports (e.g. CCSP 2009).

Although S08 stated that they used "new observational estimates," and criticized Douglass et al (2007) for their use of older datasets, S08's own analysis was based on observations ending in 1999, even though data were available to them up to the end of 2007. We show herein that with the addition of a decade of observational data, key S08 claims do not hold up, including the important comparison involving the RSS-HADCRUTv2 lapse rate.

**Methodology**

S08 claimed that there were no statistically significant differences between the observed trends in the tropical T2LT and T2 levels of the two major satellite data sets: UAH (Spencer and Christy 1990, Christy et al. 2007) and RSS (Mears et al 2003, Mears and Wentz 2005) and the corresponding ensemble means from their 49-run ensemble of models. Their test was carried out using data ending in 1999. (All discussions in this article are restricted to *tropical* data.)

S08 used a modified t-test (their $d_1^*$) to test the statistical significance of the difference between the model ensemble trend, denoted $<<b_m>>$, and the observed trend, denoted $b_0$, defined in their equation (12) as follows:

$$d_1^* = \frac{<<b_m>> - b_0}{\sqrt{\frac{1}{n_m} s(<b_m>)^2 + s(b_0)^2}} \tag{1}$$

where $s(b_0)^2$ is the estimated variance of $b_0$, $s(<b_m>)^2$ is the square of the 'inter-model standard deviation of ensemble-mean trends' and $n_m$ is the number of models (19 in S08). Values for each quantity were given in their Table 1. In their calculation of the standard error of the observed trend, $s(b_0)$, S08 adjusted the number of degrees of freedom of the trend residuals to take account of first-order (AR1) autocorrelation, using a method described in Karl et al (2006) ("CCSP"), the method itself dating back at least to Quenouille (1952). S08 considered significance using a two-sided t-test at 10, 5 and 1% levels (see their Tables III and V). We successfully emulated S08 results in their Table III using data ending in 1999 (see Supporting Information).

S08 (see their Supplementary Information) observed that it was reasonable to extrapolate model trends for comparison with post-1999 observations, a protocol that we adopt here using our implementation of their algorithm to compare observational data ending in 1999, 2007 and June 2009 to the model ensemble trends reported in S08.

S08 also compared the model ensemble surface-minus-T2LT lapse rate trend to the corresponding observed lapse rate, using the two major satellite data sets and each of four surface data sets: one land-and-ocean series (Brohan et al 2006 - HadCRUT3v) and three sea surface temperature (SST) series (Rayner et al 2003 - HadISST; Smith and Reynolds 2005 - ERSSTv2; Smith et al 2008 - ERSSTv3). T2 comparisons were not reported.

We applied our emulation of their methodology to the comparison with updated surface data. The ERSSTv3 version used in S08 was discontinued shortly after publication of S08 and replaced by a new version (ERSSTv3b) that did not incorporate as much satellite data (Smith and Reynolds 2008). We use ERSSTv3b here. ERSSTv2 is not updated as promptly as ERSSTv3b and thus our results for this comparison include data only to April 2009.

S08 did not include a comparison over land, a comparison which we make here because of the differences between land-and-ocean and ocean-only results. We examined three major tropical land data sets: CRUTEM3v (Brohan et al. 2006), GISS with 250 km smoothing (Hansen et al 2001) and NOAA-GHCN (Peterson and Vose 1997).

**Results**

*T2LT and T2 Trends*
S08 reported that the model ensemble trend for the T2LT level was 0.215 deg C (T2: 0.199). Using data to June 2009, the trend for RSS T2LT was 0.140 deg C/decade (T2:

0.104) and for UAH T2LT was 0.051 deg C/decade (T2: 0.021) and.  The model-versus-observation trend difference was significant at the 5% level under a two-sided t-test for two of the four tests (those involving UAH) - see Table 1. The model-versus-observed trend difference for one RSS test (T2) was significant under a one-sided t-test at a 10% level.

*Trends in Surface-Troposphere Lapse Rates*
S08 observed that trend tests involving surface-T2LT difference series are more "stringent" than simple trend tests, because differencing removes much of the common variability, thereby reducing AR1 autocorrelation in the trend residuals.

*Combined Land-and-Ocean:*  S08 reported that the model ensemble trend for the surface-T2LT difference series over land-and-ocean was -0.069 deg C/decade (T2: -0.053 ) i.e. troposphere warming more than surface.  For their comparison of observations over land-and-ocean, S08 compared satellite data to HADCRUT3v surface data. Using data to June 2009 (see Table 2), all four lapse rate trends exhibit statistically significant differences from models under a two-sided t-test at a 5% level (at a 1% level in 3 of the 4 cases). Thus an additional decade of data resulted in a substantial reversal of results reported in S08 for RSS comparisons.

*Ocean:* S08 reported that the model ensemble trend for the surface-T2LT difference series over ocean was -0.085 deg C/decade. Observed trends in surface-T2LT difference series involving RSS range from -0.058 to -0.083 deg C/decade, with the differences from the model ensemble mean not being significant  (see Table 3) in line with reported S08 reports.  The corresponding trends involving UAH data were positive and all differences with models were statistically significant (again in line with S08). (Corresponding results for T2 data were similar to T2LT results.)

*Land:* S08 did not report a model ensemble mean for the surface-T2LT difference series over land; the implied value from the reported land-and-ocean and ocean values was -0.012 deg C/decade. Observed trends in surface-T2LT difference series involving RSS range from +0.015 to +0.025 deg C/decade, having an opposite sign to the model ensemble, with the differences being significant at a 5% level under a two-sided t-test for comparisons to CRUTEM3 and NOAA-GHCN surface data (see Table 4). Differences with UAH satellite data were even greater.

**Discussion and Conclusions**

S08 stated that "when the RSS-derived tropospheric temperature trend is compared with four different observed estimates of surface temperature change, the surface warming is invariably amplified in the tropical troposphere, consistent with model results."  Using updated data, this statement is no longer true for the combined land-and-ocean comparison (using HADCRUT3v). While the statement continues to hold with updated data for the three SST data sets used in S08, it does not hold for the three major land data sets.

S08 stated that "even" for UAH data, observed tropical lapse rate trends are not significantly different from models and that the observed UAH tropical tropospheric trends were not significantly different from the ensemble mean. Again, this does not hold true using updated data. In this case, by re-doing the tests herein with data ending in 2007 we can show that the change from non-significance to significance does not result from end-point effects associated with 2008 temperatures. Instead it results almost entirely from a nearly 50% increase in the number of observations, thereby materially increasing the number of degrees of freedom in the statistical calculation even with autocorrelation.

Overall, the conclusion of S08 that "there is no longer a serious discrepancy between modeled and observed trends in tropical lapse rates" must be reconsidered in light of up-to-date data. The "potential inconsistency" between models and observations in the tropical region, as reported by Karl et al (2006), remains an issue.

The results reported here only pertain to an AR1 autocorrelation model as used in S08. Consideration of the impact of other autocorrelation specifications or other sources of error (e.g. Thorne et al. 2007) may well be of interest (both for this and other studies) and might contribute to a resolution of the "potential inconsistency" that concerned Karl et al (2006). However even if a reconciliation proves possible on alternative grounds, that does not affect the specific results presented here, namely that key S08 results change based on use of updated data.

| Satellite | Ensemble Trend | Observed Trend | $d_1^*$ up to 1999:12 (from S08) | $d_1^*$ up to 2009:6 (t-test percentile) | Significance under two-sided t-test (one-sided) |
|---|---|---|---|---|---|
| RSS T2LT | 0.215 | 0.140 | 0.37 | 1.08 (85.5) | - (-) |
| UAH T2LT | 0.215 | 0.051 | 1.11 | 2.42 (98.8) | ** (**) |
| RSS T2 | 0.199 | 0.104 | 0.44 | 1.41 (91.4) | - (*) |
| UAH T2 | 0.199 | 0.021 | 1.19 | 2.72 (99.4) | ** (***) |

**TABLE 1: T2LT and T2 Trend Comparisons.** Trends in deg C/decade. One, two, and three asterisks indicate model-versus-observed trend differences that are significant at the 10, 5, and 1% levels respectively.

| Surface | Satellite | Ensemble Lapse Trend | Observed Lapse Trend | $d_1^*$ up to 1999:12 (from S08) | $d_1^*$ up to 2009:6 (t-test percentile) | Significance under two-sided t-test (one-sided |
|---|---|---|---|---|---|---|
| HadCRUT3v | RSS T2LT | -0.069 | -0.028 | -0.67 | -2.34 ( 1.1) | ** (**) |
| HadCRUT3v | UAH T2LT | -0.069 | +0.061 | -3.50 | -7.00 (0.0) | *** (***) |
| HadCRUT3v | RSS T2 | -0.053 | +0.008 | n.r. | -3.30 (0.0) | *** (***) |
| HadCRUT3v | UAH T2 | -0.053 | +0.091 | n.r. | -7.73 (0.0) | *** (***) |

**TABLE 2: Lapse Rates (Land-and-Ocean).** Trends in deg C/decade. One, two, and three asterisks indicate model-versus-observed trend differences that are significant at the 10, 5, and 1% levels respectively; (two-tailed tests). All UAH lapse trends are strongly significant. nr: not reported . S08 showed absolute value of $d_1^*$.

| Surface | Satellite | Surface-minus-Satellite | | $d_1^*$ up to 1999:12 from S08 | $d_1^*$ up to 2009:6 (t-test percentile) | Significance under two-sided t-test (one-sided) |
|---|---|---|---|---|---|---|
| | | **Ensemble** | **Observed** | | | |
| HadISST1 | RSS T2LT | -0.085 | -0.083 | -0.75 | -0.07 (47.1) | - (-) |
| ERSST-v2 | RSS T2LT | -0.085 | -0.061 | -0.48 | -1.06 (14.8) | - (-) |
| ERSST-v3 | RSS T2LT | -0.085 | -0.058 | -0.12 | -1.22 (11.5) | - (-) |
| HadISST1 | UAH T2LT | -0.085 | +0.018 | -3.52 | -4.29 (0.0) | *** (***) |
| ERSST-v2 | UAH T2LT | -0.085 | +0.039 | -3.04 | -5.04 (0.0) | *** (***) |
| ERSST-v3 | UAH T2LT | -0.085 | +0.044 | -2.68 | -5.25 (0.0) | *** (***) |

**TABLE 3: Lapse Rates (Ocean).** Trends in deg C/decade. One, two, and three asterisks indicate model-versus-observed trend differences that are significant at the 10, 5, and 1% levels respectively; (two-tailed tests). S08 showed absolute value of $d_1^*$.

| Surface | Satellite | Surface-minus-T2LT | | $d_1^*$ up to 1999:12 from S08 | $d_1^*$ up to 2009:6 (t-test percentile) | Significance under two-sided t-test (one-sided) |
|---|---|---|---|---|---|---|
| | | **Ensemble** | **Observed** | | | |
| CRUTEM3 | RSS T2LT | -0.012 | +0.039 | Nr | -2.72 (0.5) | ***(***) |
| GISS-250 | RSS T2LT | -0.012 | +0.015 | Nr | -1.46 (7.5) | - (*) |
| NOAA | RSS T2LT | -0.012 | +0.025. | Nr | -2.08 (2.1) | ** (**) |
| CRUTEM3 | UAH T2LT | -0.012 | +0.104 | Nr | -6.66 (0.0) | *** (***) |
| GISS-250 | UAH T2LT | -0.012 | +0.079 | Nr | -4.82 (0.0) | *** (***) |
| NOAA | UAH T2LT | -0.012 | +0.089 | Nr | -6.21 (0.0) | *** (***) |

**TABLE 4: Lapse Rates (Land )** One, two, and three asterisks indicate model-versus-observed trend differences that are significant at the 10, 5, and 1% levels respectively; (two-tailed tests).